\documentclass[aps,pra,twocolumn,floatfix]{revtex4}

\usepackage{graphicx}
\usepackage{epsfig}
\usepackage{pst-plot}
\usepackage{bm}
\usepackage{longtable}

\usepackage{listings}
\usepackage{verbatim}
\usepackage{amsmath}

\usepackage{alg}
\usepackage{JOn}
\begin{document}

\title{MontePython: Implementing Quantum Monte Carlo using Python}
\author{J.K. Nilsen$^{1,2}$}

\affiliation{$^1$ USIT, Postboks 1059 Blindern, N-0316 Oslo, Norway\\
             $^2$ Department of Physics, University of Oslo, N-0316
             Oslo, Norway}

\begin{abstract}
 We present a cross-language C++/Python program for simulations of
quantum mechanical systems with the use of Quantum Monte Carlo (QMC)
methods. We describe a system for which to apply QMC, the algorithms
of variational Monte Carlo and diffusion Monte Carlo and we describe
how to implement theses methods in pure C++ and
C++/Python. Furthermore we
check the efficiency of the implementations in serial and parallel
cases to show that the overhead using Python can be negligible.
\end{abstract}

\maketitle

\section{Introduction}

In scientific programming there has always been a struggle between
computational efficiency and programming efficiency. On one hand, we
want a program to go as fast as possible, resorting to low-level
programming languages like FORTRAN77 and C which can be difficult to read
and even harder to debug. On the other hand, we want the
programming process to be as efficient as possible, turning to
high-level software like Matlab, Octave, Maple, R and S+. Here
features like clean 
syntax, interactive command execution, integrated simulation and
visualization and rich documentation make us
feel more productive.  However, if we have some well tested and fast
routines written in low-level language, interfacing these routines
with, e.g., Matlab is rather cumbersome. Most often, we will end up
using similar Matlab routines, which are often written as generally as
possible at the cost of computing efficiency.

Recently, the programming language Python \cite{python} has emerged as
a potential competitor to Matlab. Python is a very powerful
programming language which, when extended with numerical and visual
modules like SciPy \cite{scipy}, shares many of the features of
Matlab. In addition, Python was designed to be extendible with
compiled code for efficiency and several tools are available for doing
so.

In this paper we will demonstrate how Python can be extended with
compiled code to yield an efficient scientific program. Specifically,
we will start with a Monte Carlo simulator written in C++ and, with
the help of SWIG \cite{swig}, reuse the C++ code in a Python Monte
Carlo simulator. We will show that this porting from low-level to
high-level code can be achieved without significant loss of efficiency.

The remainder of this paper is organized as follows. In section
\ref{sect:system} we define the system we apply the Monte Carlo
simulator to. Section \ref{sect:algorithms} discuss in some detail the
algorithms we are going to use, i.e., variational Monte Carlo and
diffusion Monte Carlo. Next, we go through the implementations of
diffusion Monte Carlo, both in C++ and Python, in section
\ref{sect:algorithms}. Furthermore, section \ref{sect:benchmarks}
compare the efficiency of C++ and Python for varying numbers of CPUs
and section \ref{sect:visualizing} visualize the output from
diffusion Monte Carlo with the use of Python. Finally, we round off
with some remarks in section \ref{sect:conclusion}.

\section{The system}
\label{sect:system}

Quantum Monte Carlo (QMC) has a wide range of
applications, for example
studies of Bose-Einstein condensates of dilute atomic gases (bosonic
systems) \cite{dubois2003} and studies of so-called quantum dots
(fermionic systems) \cite{harju2005},
electrons confined between layers in semi-conductors. In this paper we
will focus on a model which is meant to reproduce the results from an
experiment by 
Anderson \& al. \cite{anderson}. Anderson \& al. cooled down $4\times
10^6$ $^{87}$Rb to 
temperatures in the order of $100\, nK$ to observe Bose-Einstein
condensation in the dilute gas. Our physical motivation in this paper
is to model numerically this fascinating experiment. This
should be done in an as general as possible way, so that we can expand
our computations to systems not yet explored in experiments. We will
in this section go through the steps needed to put the experiment into
the framework of QMC. 

In QMC the goal is to solve the Schr\"odinger equation
\begin{equation}
i\hbar\frac{\partial}{\partial t}\Psi(\vec R,t) = H\Psi(\vec R,t),
\label{eq:SE_time}
\end{equation}
or rather the time independent version
\begin{equation}
H\Psi(\vec R) = E\Psi(\vec R).
\label{eq:SE}
\end{equation}
Thus, to model the experiment above using Quantum Monte Carlo methods, all we
need is a Hamiltonian and a trial wave function. 
The Hamiltonian for $N$ trapped interacting atoms is given by
\begin{equation}
H=-\frac{\hbar^2}{2m}\sum_{i=1}^{N}\nabla_i^2 + \sum_{i=1}^{N}
V_{ext}(\vec r_i) +\sum_{i<j}^{N} V_{int}(|\vec r_i - \vec r_j|).
\label{eq:Hamiltonian}
\end{equation}
Taking advantage of the fact that the gas is dilute, we can describe
the two-body interaction $V_{int}(|\vec r_i - \vec r_j|)$ by a
hard-core potential of radius $a$, where $a$ is the scattering length,
thus treating the atoms as hard spheres \cite{nilsen2005}.

We define the trial wave function by
\begin{equation}
\Psi_T = \prod_i g(\vec r_i)\prod_{i<j} f(r_{ij}),
\label{eq:wf}
\end{equation}
where $g(\vec r_i)$ describes the interaction between one particle and
the external potential, $V_{ext}$, while the two-body correlation
function $f(r_{ij})$ describes the interaction between two
particles. The function $f(r_{ij})$ is the solution of the
Schr\"odinger equation for a pair of atoms at very low energy
interacting via a hard-core potential of radius $a$. The ansatz for
$f(r)$ reads
\begin{equation}
f(r)=\left\{
\begin{array}{lll}
  \left(1-a/r\right)&\quad&r>a\\
  0                 &\quad&r\leq a.
\end{array}
\right.
\end{equation}
Besides being physically motivated, this type of correlation has been
successfully used in refs. \cite{dubois2001} and \cite{dubois2003} to
study both spherically symmetric and deformed traps. In the experiment
of Anderson \& al., the particles were trapped in a disk-shaped harmonic
oscillator potential. This corresponds to using an external potential
\begin{equation}
V_{ext} = \frac{m}{2}\left(\omega_\perp x^2 + \omega_\perp y^2 +
\omega_z z^2\right)
\label{eq:HO}
\end{equation}
If we neglect the particle-particle interaction and insert the
potential of eq. \refeq{eq:HO} into eq. \refeq{eq:SE} we obtain
\begin{equation}
g(\vec r)=A(\alpha)\lambda^{1/4}\exp\left(-\alpha(x^2+y^2+\lambda
z^2)\right)
\end{equation}
where $\alpha$ is taken as the variational parameter of the
calculation and $A(\alpha)=(2\alpha/\pi)^{3/4}$ is a normalization
constant. The parameter $\lambda=\omega_z/\omega_\perp$ is kept
constant and set equal to the asymmetry of the trap. Still following
Anderson \& al. we let $\lambda=\sqrt{8}$ throughout this paper.

\section{The algorithms}
\label{sect:algorithms}

In this section we will go through the algorithms used in this
paper. First, we will discuss Monte Carlo integration in
general. Second, we state the Variational Monte Carlo algorithm and
last, we go through the Diffusion Monte Carlo algorithm.

\subsection{Monte Carlo Integration}

Monte Carlo integration is best described through conventional
numerical integration methods. In conventional methods we fix the
evaluation points and weights of the integrand in advance,
\begin{equation}
\int_\Omega f(\vec r) d\Omega = \sum_{i=1}^m \omega_i f(r_i) + \mathcal O\left(\frac{1}{m}\right).
\end{equation}
If we choose the evaluation points with equal spacing over the
integration area, a two dimensional integral can for example be
written as
\begin{equation}
\int_0^1\int_0^1 f(x,y)\,dxdy = \frac{1}{m^2}\sum_{i=1}^m
\sum_{i=1}^mf(x_i,y_j) \mathcal O\left(\frac{1}{m}\right).
\end{equation}
For $N$ dimensions we have to carry out $m^N$ evaluations of
$f(r)$ to obtain an accuracy of $\mathcal
O\left(\frac{1}{m}\right)$. In Monte Carlo integration the integrand
is evaluated at random points $\vec r_i$ drawn from an arbitrary
probability distribution $\rho(\vec r)$,
\begin{equation}
\int_\Omega f(\vec r)\,d\Omega = \int_\Omega g(\vec r)\rho(\vec
r)\,d\Omega = \sum_{i=1}^mg(r_i) + \mathcal
O\left(\frac{1}{\sqrt{m}}\right).
\end{equation}
The main advantage of this scheme is that it is independent of the
number of dimensions of $\vec R$. Note, however, that the efficiency
of the integration depends on a good choice for $\rho(\vec R)$.

\subsubsection{Metropolis Algorithm}

The Metropolis Algorithm \cite{metropolis} generates a stochastic
sequence of phase space points that samples a given probability
distribution. In Quantum Monte Carlo methods each point in phase space
represents a vector $\vec R = \{\vec r_1,\vec r_2,\ldots,\vec r_N\}$
in Hilbert space. Here $\vec r_i$ represents all degrees of freedom
for particle $i$. Coupled with a quantum mechanical operator (like the
Hamiltonian in eq. \refeq{eq:Hamiltonian}) each such point can be
associated with physical quantities (the Hamiltonian gives the energy
of the system). The fundamental idea behind the Metropolis algorithm 
is that the sequence of individual {\it samples} of these quantities
can be combined to arrive at average values which describes the
quantum mechanical state of the system. The Metropolis algorithm
provides the sample points. From an initial position in phase space a
{\it proposed move} is generated and the move is either {\it accepted}
or {\it rejected} according to the Metropolis algorithm
\begin{equation}
P_A(\vec R',\vec R) = min\left(1,\frac{\rho(\vec R')}{\rho(\vec R)}\right).
\label{eq:metropolis}
\end{equation}
where $P_A(\vec R',\vec R)$ is the probability of moving the particle
from $\vec R$ to $\vec R'$. In this way a {\it random walk} generates
a sequence $\{\vec R_0,\vec R_1,\ldots,\vec R_i,\ldots\}$ of points in
phase space. It is important that all points must be accessible from
any starting point; the random walk must be {\it ergodic}.

Metropolis \& al. showed that the sampling is most easily
achieved if the points $\vec R$ form a {\it Markov chain}. A random
walk is Markovian if each point in the chain depends only on the
position of the previous point.

\subsection{Variational Monte Carlo}
\label{sect:vmc}

Variational Monte Carlo (VMC) is the starting point of all Monte Carlo
calculations in that we need an optimized trial wave function as input
to the other Monte Carlo methods. In VMC we combine Monte Carlo
integration with the Metropolis algorithm and the variational
principle to get the trial wave function that yields the lowest
energy.

The {\it variational principle} states that, given a variational wave
function $\psi_\alpha$ (where $\alpha=\alpha_1,\alpha_2,\ldots$ denote
a set of variational parameters), the energy expectation value of
$\psi_\alpha$ provides an upper bound to the ground state energy,
i.e.,
\begin{equation}
\langle \oper H\rangle_\alpha=\frac{\int\psi_\alpha^*(\vec R) \oper
  H\psi_\alpha(\vec R)\,d\vec R}{\int\psi_\alpha^(\vec
  R)*\psi_\alpha(\vec R)\,d\vec R} \geq
\langle\oper H\rangle.
\label{eq:variational_principle}
\end{equation}

Rewriting eq. \refeq{eq:variational_principle},
\begin{equation}
\langle\oper H\rangle_\alpha = \int E_L(\vec R)\frac{|\psi_\alpha(\vec R)|^2}{\int
  |\psi_\alpha(\vec R)|^2\,d\vec R}\,d\vec R
\end{equation}
with the local energy
\begin{equation}
E_L = \frac{1}{\psi_\alpha(\vec R)}\oper H\psi_\alpha(\vec R)
\label{eq:local_energy}
\end{equation}
we can interpret the square of the wave function divided by its norm
as the probability distribution of the system, arriving at
\begin{equation}
\langle\oper H\rangle_\alpha = \int E_L(\vec R)\rho(\vec R)\,d\vec R
\label{eq:vmc_integral}
\end{equation}
where
\begin{equation}
\rho(\vec R) = \frac{|\psi_\alpha(\vec R)|^2}{\int |\psi_\alpha(\vec R)|^2\,d\vec R}.
\end{equation}
We then carry out the integral of eq. \refeq{eq:vmc_integral} with Monte Carlo
integration. We move a walker randomly through phase space according
to the Metropolis algorithm, and sample the local energy with each
move. This way we get a statistical evaluation of the integral.

The VMC algorithm consists of two distinct parts. First, as the Metropolis
algorithm reproduces the probability distribution in the limit
\matenv{t\to\infty}, you need a {\it thermalization}, where the
walker propagates according to the Metropolis algorithm in order to
equilibrate it to the probability distribution \matenv{\rho(\vec
  R)}. Second, you continue to move the walker, but you sample
energies and other observables for computation of averages and other
statistical observables.

\begin{algorithm}[t]
  \caption{VMC algorithm}
  \label{alg:vmc}
  {\it Generate initial randomized configuration}\\
  
  \algforto{0}{VMC steps}
  \begin{tabular}{l}
    \algforto{0}{particles}
    \begin{tabular}{l}
      {\it Propose move} \matenv{\vec r\to\vec r'}\\
      {\it Compute ratio} \matenv{w=|\psi(\vec R')/\psi(\vec R)|^2}\\
      {\it Accept or reject according to the Metropolis probability}
      \matenv{min(1,w)}\\
    \end{tabular}\\
    {\it Sample the contributions to the local energy and other
    observables in this move}\\
  \end{tabular}\\
  {\it Sample the contributions to the local energy and other
  observables for this series of movements}\\
  {\it Do statistics and variate the parameters in wave function
  according to the statistics}\\
\end{algorithm}

In algorithm \ref{alg:vmc} the particles are moved one by one and not
as a whole configuration\footnote{The observables are, however, only
  calculated after all particles have been moved.}. This improves the
efficiency of the algorithm for larger systems, as moving the whole
configuration requires decreasing the steps to maintain the acceptance
ratio \cite{ab_initio}.

\subsection{Diffusion Monte Carlo}
\label{sect:dmc}

In Diffusion Monte Carlo (DMC) we seek to solve the Schr\"odinger
equation in imaginary 
time. This involves Monte Carlo
integration of a Green's function. As the Green's function is approximated by
splitting it up in a diffusional part (which has the form of a
Gaussian) and a branching part we also need a Gaussian random
generator and a way to create and destroy walkers.

\subsubsection{Basic Ideas of DMC}

The basic ingredients of DMC are \cite{guardiola}:

\begin{enumerate}
\item It considers the Schr\"odinger equation in imaginary time,
\begin{equation}
-\frac{\partial d\psi(\vec R,t)}{\partial t} = [\oper H-E]\psi(\vec
R,t),
\label{eq:SL_imaginary_time}
\end{equation}
where \matenv{\vec R} represents the set of all coordinates. The
formal solution of \refeq{eq:SL_imaginary_time} is
\begin{equation}
\psi(\vec R,t) = e^{-[H-E]t}\psi(\vec R,0),
\end{equation}
where \matenv{\exp[-(H-E)t]} is called the {\it Green's function}, and
\matenv E is a convenient energy shift.
\item The wave function is positive definite everywhere, as it happens
  with the ground state of a bosonic system, so it may be considered
  as a probability distribution function. (This assumption leads to
  difficulties when we consider fermionic systems, where the wave
  functions are anti-symmetric and special care needs to be made.)
\item The wave function is represented by a set of random vectors
  \matenv{\{R_1,R_2,\ldots,R_M\}}, in such a form that the time
  evolution of the wave function is actually represented by the
  evolution of the set of walkers.
\item The actual computation of the time evolution is done in small
  time steps \matenv\tau, and the Green's function is approximated
  accordingly,
\begin{equation}
e^{-[H-E]t}=\prod_{i=1}^ne^{-[H-E]\tau},
\end{equation}
where \matenv{\tau=t/n}.
\item The imaginary time evolution of an arbitrary starting state
  \matenv{\psi(\vec R,0)}, once expanded in the basis of stationary
  states of the Hamilton operator
\begin{equation}
\psi(\vec R,0)=\sum_\nu C_\nu\phi_nu(\vec R)
\end{equation}
is given by
\begin{equation}
\psi(\vec R,t)=\sum_\nu e^{-[E_\nu-E]t}C_\nu\phi_\nu(\vec R),
\label{eq:imaginary_time_evolution}
\end{equation}
in such a way that the lowest energy components will have the largest
  amplitudes after a long elapsed time, and in the \matenv{t\to\infty}
  limit the most important amplitude will correspond to the ground
  state (if \matenv{C_0\neq0})\footnote{This can easily be seen by
    replacing \matenv E with the ground state energy \matenv{E_0} in
    eq. \refeq{eq:imaginary_time_evolution}. As \matenv{E_0} is the
    lowest energy, we will get
    \matenv{\lim_{t\to\infty}\sum_\nu\exp[-(E_\nu-E_0)t]
      \phi_\nu = C_0\phi_0}.}.
\item An improvement of this scheme is the introduction of
  {\it importance sampling}.
\end{enumerate}

The scheme is quite simple; once we have found an appropriate
approximation for the short-time Green's function and determined a
starting state, the job consists in representing the starting state by
a collection of walkers and letting them evolve in time, i.e.,
obtaining a collection of walkers from the old collection of walkers,
up to a time large enough so that all other states than the ground
state are negligible.

\subsubsection{Shortime Green's Function}

In coordinate representation the Green's function is given by the matrix
element
\begin{equation}
G(\vec R',\vec R,t) = \bra{\vec R'}e^{-[H-E]t}\ket{\vec R},
\end{equation}
and the time evolution equation is
\begin{equation}
\psi(\vec R,t) = \int G(\vec R', \vec R, t)\psi(\vec R,0)\,d\vec R.
\end{equation}
At \matenv{t=0} the value of the Green's function is
\begin{equation}
G(\vec R',\vec R,0) = \delta(\vec R'-\vec R).
\label{eq:green_boundary_condition}
\end{equation}
From the operatorial representation of the Green's function we can
easily obtain the formal differential equation
\begin{equation}
-\frac{\partial G}{\partial t} = [\oper H - E]G,
\end{equation}
or written out in the coordinate representation,
\begin{equation}
\begin{split}
-\frac{\partial G(\vec R',\vec R,t)}{\partial t} =
&\left[-\frac{\hbar^2}{2m}\sum_i\nabla_i^2+\sum_{i<j}V(r_ {ij}) -
  E\right]\times\\
&\quad G(\vec R',\vec R,t)
\end{split}
\end{equation}
with the boundary condition \refeq{eq:green_boundary_condition}.

The Hamiltonian, eq. \refeq{eq:Hamiltonian}, can be rewritten as
\begin{equation}
\oper H = -D\nabla^2 + V(\vec R) = \oper K + \oper V, 
\end{equation}
where $D=\hbar/2m$ is a diffusion constant, $\oper K$ is the kinetic
energy operator 
and $\oper V$ is the potential energy operator. The Green's function can
now be written as $G(\vec R',\vec R,t) = \bra{\vec R'}e^{-[\oper
    K+\oper V-E]t}\ket{\vec R}$.
The main problem in obtaining the Green's function is that $\oper K$ and
$\oper V$ does not commute. The Green's functions related exclusively to
the kinetic operator and the potential operator is, however, readily
determined.

The kinetic operator can be written as
\begin{equation}
G_K(\vec R',\vec R,t) = \frac{1}{(2\pi)^{3N}} \int e^{-i\vec k\vec
  R'}e^{-Dtk^2}e^{i\vec k\vec
  R}\,d\vec k,
\label{eq:kinetic_greensfunction}
\end{equation}
If we carry out the integral of eq. \refeq{eq:kinetic_greensfunction} we
get the Green's function
\begin{equation}
G_K(\vec R', \vec R,t) = \frac{1}{(4\pi Dt)^{3N/2}}e^{-(\vec R-\vec
  R')^2/4Dt}.
\end{equation}
It is easily checked that this Green's function is the Dirac delta
function \matenv{\delta(\vec R - \vec R')} in the \matenv{t=0} limit.

The Green's function related to the potential is even simpler, for
momentum independent interactions. The potential is then a
{\it local operator} and the Green's function
\begin{equation}
G_V(\vec R',\vec R,t) = e^{-V(\vec R)t}\delta(\vec R - \vec R').
\end{equation}

Approximate forms for the full Green's function are obtained by using
the following approximations (with time step \matenv\tau) \cite{guardiola}:
\begin{equation}
\begin{split}
\bra{\vec R'}e^{-(K+V)\tau}\ket{\vec R} &= \bra{\vec
  R'}e^{-K\tau}\ket{\vec R''}\bra{\vec R''}e^{-V\tau}\ket{\vec R} +
  \mathcal O(\tau^2)\\
&= \bra{\vec R'}e^{-V\tau}\ket{\vec R''}\bra{\vec R''}e^{-K\tau}\ket{\vec
  R} + \mathcal O(\tau^2)\\
= &\bra{\vec R'}e^{-V\tau/2}\ket{\vec R''}\bra{\vec R''}e^{-K\tau}\ket{\vec
  R'''}\times\\
  &\quad \bra{\vec R'''}e^{-V\tau/2}\ket{\vec R} + \mathcal O(\tau^3).
\end{split}
\end{equation}
In all cases an integration over the internal coordinates \matenv{\vec
  R''} and \matenv{\vec R'''} must be understood, being easily carried out
by using the delta functions appearing in \matenv{G_V}. The practical
equations are
\begin{equation}
\begin{split}
G(\vec R',\vec R,\tau) &= G_K(\vec R',\vec R,\tau)e^{[E-V(\vec R)]\tau} +
\mathcal O(\tau^2)\\
&= e^{[E-V(\vec R')]\tau}G_K(\vec R',\vec R,\tau) + \mathcal O(\tau^2)\\
&= e^{[E-\{V(\vec R')+V(\vec R)\}/2]\tau}G_K(\vec R',\vec R,\tau) +
\mathcal O(\tau^3).
\end{split}
\label{eq:various_green_functions}
\end{equation}

\subsubsection{Importance Sampling}

An important improvement to the DMC scheme above is, as mentioned
above, the use of {\it importance sampling}. In problems with
singularities in the potential (e.g., the Coulomb potential) the Green's
function $\exp[-(H-E)t]$ will reach unbounded values, leading to an
unstable algorithm. Even without singularities the scheme above is
inefficient. This is due to the fact that we have imposed no
restrictions as to where the walkers will walk. 

Consider the
imaginary time evolution of the quantity
\begin{equation}
f(\vec R, t) = \psi_T(\vec R)\psi(\vec R,t),
\end{equation}
where \matenv\psi~is the wave function satisfying the Schr\"odinger
equation and \matenv{\psi_T} is a time independent trial function,
preferably close to the exact ground state wave function. It is not
necessarily the starting wave function, even if it normally also is
taken to be the starting wave function.

The time evolution of \matenv{f(\vec R,t)} is given by inserting
\matenv{\psi_T} in the Schr\"odinger equation:
\begin{equation}
\begin{split}
-\frac{\partial \psi(\vec R,t)}{\partial t}\psi_T(\vec R) = -
&D(\nabla^2\psi(\vec R,t))\psi_T(\vec R) +\\ 
&(V-E)\psi(\vec R,t)\psi_T(\vec R)
\end{split}
\end{equation}
with \matenv{D=\hbar^2/2m}. As
\begin{equation}
\nabla^2(\psi_T\psi) = (\nabla^2\psi)\psi_T + \psi(\nabla^2\psi_T) +
2(\nabla\psi)(\nabla\psi_T)
\end{equation}
we have
\begin{equation}
-\frac{\partial f}{\partial t} = -D\nabla^2f +
2D(\nabla\psi)(\nabla\psi_T) + D\psi(\nabla^2\psi_T) + (V-E)f.
\end{equation}
Introducing the so called {\it drift force} (which is a drift
velocity), given by
\begin{equation}
\vec F=\frac{2}{\psi_T}\nabla\psi_T,
\end{equation}
and using that the local energy is
\begin{equation}
E_L = -D\frac{\nabla^2\psi_T}{\psi_T}+V
\end{equation}
we end up with the (imaginary) time evolution of \matenv f as
\begin{equation}
-\frac{\partial f}{\partial t} = -D\nabla^2f + D\nabla\cdot(\vec Ff)+
(E_L-E)f.
\label{eq:importance_time_evolution}
\end{equation}
The transformed Hamilton operator working on \matenv f in
eq. \eqref{eq:importance_time_evolution} may be written as a sum of
three terms
\begin{equation}
\begin{split}
\oper H &= \oper K + \oper F + \oper L\\
\oper K &=  -D\nabla^2\\
\oper F &= -D(\nabla\cdot\vec F(\vec R)) + \vec F(\vec R)\cdot\nabla\\
\oper L &= E_L(\vec R)
\end{split}
\end{equation}
corresponding respectively to the kinetic part, the drift part and the
local energy part.

An \matenv{\mathcal O(\tau^2)} approximation of the Green's function is
given by \cite{reynolds1982}:
\begin{equation}
\begin{split}
\bra{\vec R'}G\ket{\vec R} = &\frac{1}{(4\pi D\tau)^{3N/2}}e^{-[\vec R'
    - \vec R - D\tau\vec F(\vec R)]^2/4D\tau}\times\\
&\quad e^{E\tau - [E_L(\vec R') + E_L(\vec R)]\tau/2} + \mathcal O(\tau^2).
\end{split}
\label{eq:explicit_green_ot2}
\end{equation}  
while an \matenv{\mathcal O(\tau^3)} approximation the Green's function is
obtained from \cite{sarsa2001}
\begin{equation}
G =
e^{E\tau}e^{-L/2\tau}e^{-F/2\tau}e^{-K\tau}e^{-F/2\tau}e^{-L/2\tau} +
\mathcal O(\tau^3). 
\label{eq:explicit_green_ot3}
\end{equation}
 
\subsubsection{DMC Algorithm}

In algorithm \ref{alg:dmc} we state the DMC algorithm corresponding to
eq. \refeq{eq:explicit_green_ot2}. The algorithm corresponding to
eq. \refeq{eq:explicit_green_ot3} is similar except that the move is split
into four parts due to the splitting of the drift operator. $\xi$ in
the move part of algorithm \ref{alg:dmc} is drawn from the
multivariate Gaussian distribution with null mean and
$\sigma=\sqrt{2D\tau}$, the solution of the kinetic Green's function.

\begin{algorithm}[t]
  \caption{DMC Algorithm}
  \label{alg:dmc}
  Generate an initial set of random walkers with the Metropolis Algorithm\\
  \algforto{0}{time}
  \begin{tabular}{l}
    \algforto{0}{$N_{walkers}$}
    \begin{tabular}{l}
      Diffusion:\\
      \algforto{0}{particles}
      \begin{tabular}{l}
	propose move \matenv{\vec r'=\vec r + D\tau\vec F(\vec
	  r) + \xi}\\
      \end{tabular}\\
      Branching; calculate replication factor:\\
      \matenv{n=int(\exp\{\tau(E_L(\vec R)/2+E_L(\vec R')/2-E)\})}\\
      \algif{\matenv{n=0}}
      \begin{tabular}{l}
	Kill the walker\\
      \end{tabular}\\
      \algif{\matenv{n>0}}
      \begin{tabular}{l}
	Allow the walker to make \matenv{n-1} clones\\
      \end{tabular}\\
      Remove dead walkers, and make new clones\\
      Check walker population and adjust trial energy\\
      sample contributions to observable\\
    \end{tabular}\\
  \end{tabular}\\
\end{algorithm}

It can be showed \cite{ab_initio} that importance sampling may be
used in Variational Monte Carlo as well, using the Fokker-Planck
formalism. The resulting algorithm is identical to alg. \ref{alg:dmc},
but without the branching term. We will therefore concentrate on
diffusion Monte Carlo in the rest of this article.

\section{The implementations}
\label{sect:implementations}

In the previous sections we have identified a physical system to
simulate and found algorithms to use in the simulations. One important
question that remains is how we implement the system and the
algorithms. In this section we will propose three different
approaches. They all use the same algorithms, they solve the same
systems, with identical results, and they are all written in an object
oriented way. In fact, most of the code is the same for all three
approaches. The only difference in the implementations is the amount
of time spent in low level, compiled language (represented by C++) versus
time spent in high level, interpreted language (represented by
Python). The assumption is that compiled code is faster while
interpreted code is clearer, easier to debug and easier to expand. We
will in this section go through a pure C++ implementation, a straight
forward Python approach and a more involved Python approach.

\subsection{C++ implementation}

\begin{figure*}[t]
  \begin{center}
    \includegraphics[width=0.8\textwidth]{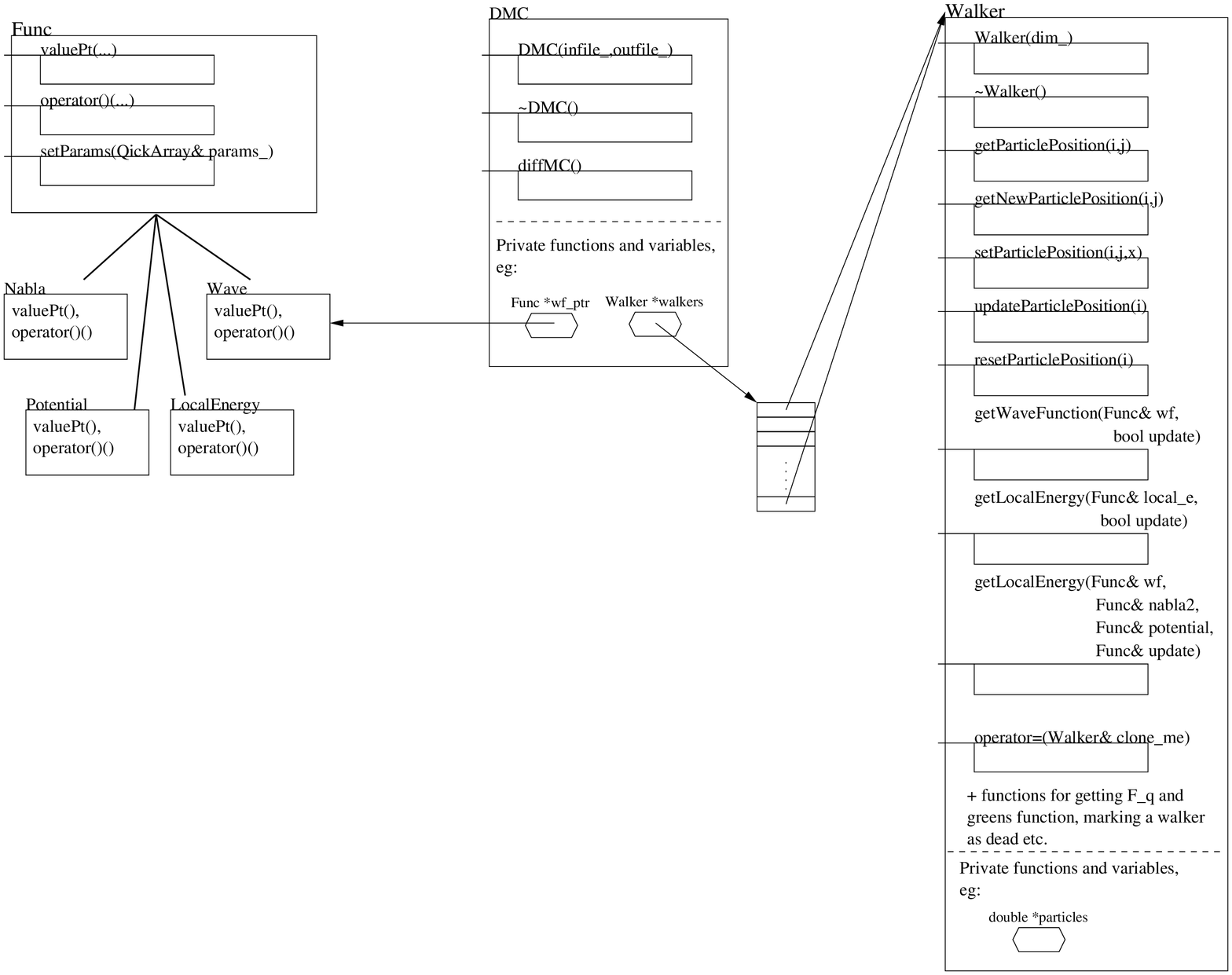}
  \end{center}
  \caption{Class diagram of DMC} 
  \label{fig:class_diagram_dmc}
\end{figure*}

\begin{figure*}
  \begin{center}
    \includegraphics[width=.7\textwidth]{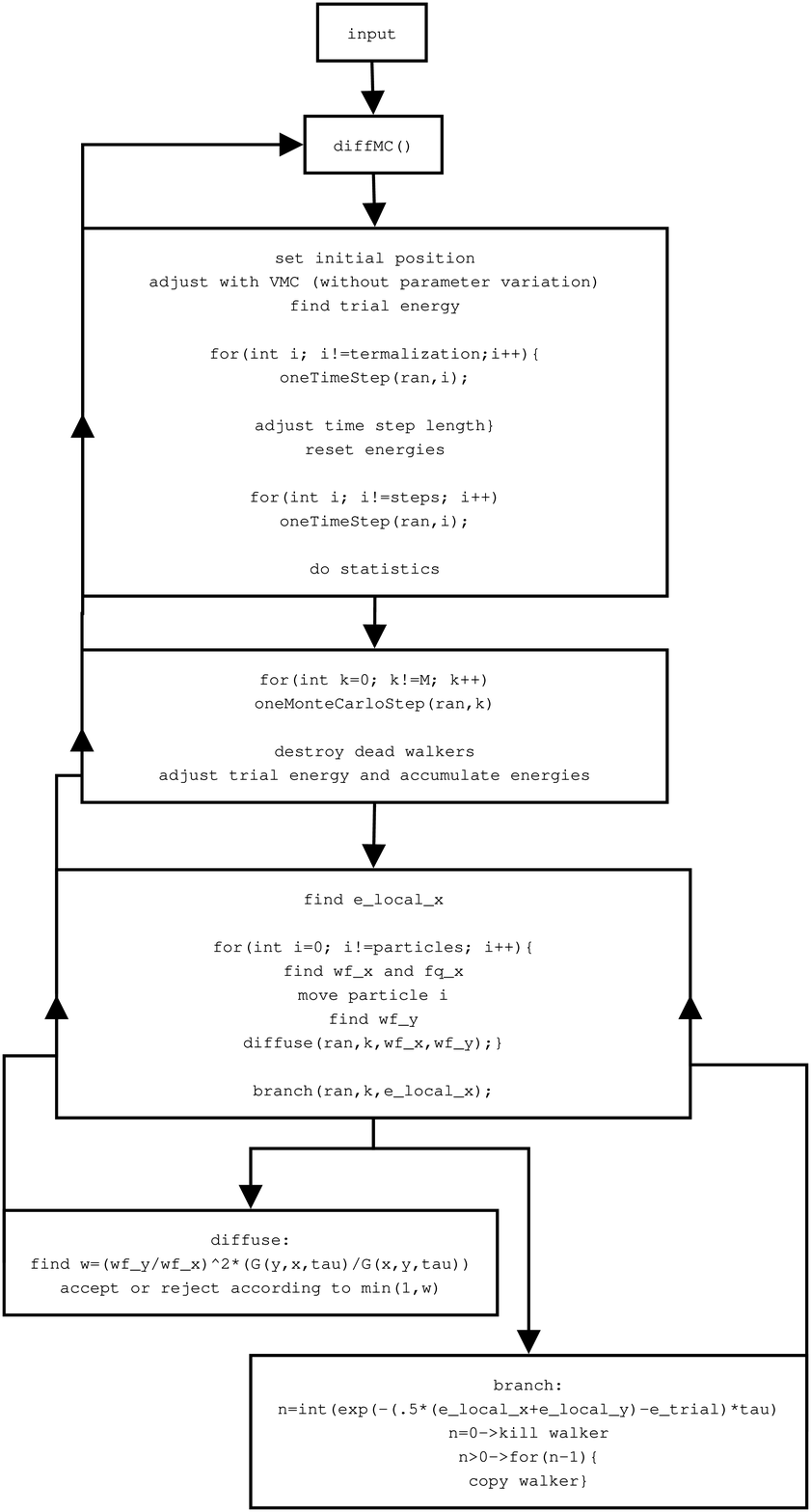}
  \end{center}
  \caption{Float diagram of DMC} 
  \label{fig:float_diagram_dmc}
\end{figure*}

The base of our implementations is a serial diffusion Monte
Carlo (DMC) solver written in C++. The Python solvers are both heavily
based on this code. We will therefore first go through the C++
implementation of DMC. In figs. \ref{fig:class_diagram_dmc} and
\ref{fig:float_diagram_dmc} we present the class diagram and float
diagram of the C++ implementation. 

In fig. \ref{fig:class_diagram_dmc} we
show three classes, class DMC, class Func and class Walker. 
\begin{itemize}
\item The class DMC contains the DMC algorithm, implemented in
the function diffMC() (and helper functions to clean up the code). It
also contains a pointer to the class Func and an array of walker
objects (or just walkers). 

\item The class Func contains functors, i.e., classes whose only purpose is
  to receive a set of numerical values and transform these to
  numerical output (not unlike mathematical functions). Specifically,
  Func contains 
different wave functions (with corresponding analytic local energies
and quantum forces if implemented) along with generic functions for
the gradient and Laplace operator. The different functions of the
  systems are subclasses derived from general functions to ensure that
  the functions of all the systems have the same input and output.

\item The class Walker contains all the
physical information of a walker, that is, its position in phase space
(and function for setting and getting the position) and functions for
getting physical values like the energy of the walker and the wave
function of the walker.
\end{itemize}

The advantage of this division of the program is quite clear. The
class DMC contains the DMC algorithm and may easily be replaced with
other Monte Carlo algorithms, like the already mentioned variational
Monte Carlo, Green's Function Monte Carlo and so on. These replacements
will neither affect the systems implemented in class Func nor the
physical information of the walkers. Likewise, new systems may be
implemented without changing the code of the
algorithm\footnote{Except, of course,
  that the DMC class has to know that the new system exists}. The wave
function and the potential (or optionally an analytic expression of the
local energy and the quantum force) are sent to the walkers as
pointers to Func objects and are as such not known to the walkers at
compile time. 

Fig. \ref{fig:float_diagram_dmc} shows the float diagram of the DMC
program. The algorithm is divided into functions so that,
e.g., the function diffMC() contains a loop calling the
function oneTimeStep(), which in turn loops over oneMonteCarloStep()
and so on. Each such function is represented by a box in the float
diagrams.

Looking at the float diagram, fig. \ref{fig:float_diagram_dmc}, it is
easy to realize that most of the time
of computation is spent in the bottom boxes of the diagram. When
implementing the DMC in Python the bottom boxes should be kept in C++
while only diffMC() (which is in broad lines the hole DMC algorithm)
will be in Python code.

\subsubsection{Checkpointing}

And aspect which is frequently forgotten when writing a scientific
program is the aspect of checkpointing. A Monte Carlo simulation may
easily take several days, or even weeks and months. This would be
unfeasible without some way to stop and start the simulation in case
of computer crashes, power losses or overeager computer managers. In
checkpointing we store all information needed to resume the
computation at given steps of the simulation. The challenge is to
identify the steps at which the checkpointing should be made. The
checkpoints should be made frequently enough to save time compared to
starting all over, but not so frequently that the simulation is
significantly slowed down.

In variational Monte
Carlo it suffices to write a new initialization file where the number
of steps is reduced to what is remaining of the original number of
steps and a random seed so that we continue the random stream we have
started. The latter is important if we want to get reproducible
results. As the amount of data stored in a checkpoint is so small, we
can do it after every step without reducing speed. However, making a
checkpoint during the movement of the particles would be quite
cumbersome and the amount of data needed to store the checkpoint would
increase dramatically.

Again, diffusion Monte Carlo is more challenging. As generating a
starting state in effect takes a variational Monte Carlo run, we have
to store all the walkers at every checkpoint. This involves storing
all particle positions, the last calculated local energy and quantum
force (which is a vector) and so on, for every walker. In the C++
implementation this is realized by the functions getBuffer and
setBuffer in the Walker objects. In a call to getBuffer the walker
puts all it's information into a character array. In a checkpoint
these arrays are concatenated and dumped to file. When restarting the
program, the arrays are read from the file and sent to the walkers
through setBuffer. The checkpoints are, as for variational Monte
Carlo, made after every time step.

\subsubsection{Generating random numbers}

When all is said and done the most important function in a Monte Carlo
method is the random number generator. The Monte Carlo integration
depends on a walker's ability to reach all points in phase space from
its starting point. If the random numbers determining the movement of
the walker are in some way correlated, the walker will lose this
ability. A good random number generator is therefore of great
importance. Consider the simple one-dimensional definite 
integral
\begin{equation}
F=\int_0^1\,f(x)dx
\label{eq:intF}
\end{equation}
To solve this equation numerically, we approximate $F$ in terms of
$F_N$:
\begin{equation}
F = \lim_{N\to\infty}\,F_N
\label{eq:numintF}
\end{equation}
where
\begin{equation}
F_N=\frac{1}{N}\sum_{i=1}^N f(x_i).
\end{equation}
When we solve eq. \refeq{eq:intF} using Monte Carlo integration, we draw
the sample points $\{x_i\}$ randomly from a given probability density
function. However, as a computer only has a finite sized set of numbers
available, we have to use random numbers generated from a
pseudo-random number generator (PRNG). For every PRNG there is a
finite number of pseudo-random numbers, known as the cycle length of the
PRNG. When this cycle length is reached eq. \refeq{eq:numintF} will
cease to converge. This may not seem like a serious problem as the
cycle length can be made quite large by using better PRNGs. However,
we have to take care to choose a good PRNG. To take an example, the
PRNG \texttt{ran0} (see \cite{numerical_recipes}) has a cycle length
of about $2.1\times 10^9$. On a 3.40GHz Intel(R) Xeon(TM) CPU
\texttt{ran0} takes 40 seconds to run through one cycle. It is obvious
that using this (widely used) PRNG will lead to problems when a
Diffusion Monte Carlo simulation takes several days of CPU time.

The generation of random numbers is a science in itself and, though of
great importance to Monte Carlo methods, we will not go through this
aspect in detail. We can advise the interested reader to read the
introduction of \cite{donald_knuth}. In the simulations in this paper
we have used a 64-bit linear congruential generator with prime addend
\cite{SPRNG,testingPRNG} which has a period of $2^{64}$. Linear
congruential generators may have correlations between numbers that are
separated by a power of 2. We should therefore take care to avoid using
this generator in batches of powers of 2. In our case, this happens
in all 3D simulations with variational Monte Carlo. In algorithm
\ref{alg:vmc} we make 3 calls to the random generator to move one
particle, then one call to  determine whether the move should be
rejected or accepted. This is repeated for every particle. In our
case, this can be avoided by letting the acceptance/rejection
procedure use another random stream than the movement and making sure
these streams are uncorrelated with each other.  

\subsection{Parallelizing the C++ implementation}

To parallelize Variational Monte Carlo (VMC) is embarrassingly
easy. As long as you ensure that all the calculations use different sets
of random numbers (and thereby ensuring that the calculations are
uncorrelated) the algorithm is parallelized by running an independent
calculation on each node. The communication between the nodes is
restricted to spreading the input parameters before the calculations
and collecting the output after calculation. The parallel efficiency
is essentially 100\%, and the calculation can theoretically use any
number of nodes without efficiency loss.

The parallelization of Diffusion Monte Carlo (DMC) is more
cumbersome. This is due to the branching part in algorithm
\ref{alg:dmc} where walkers are killed or reproduced. If we had
parallelized DMC in a straight forward way, i.e., by starting one DMC
run per node with different sets of random numbers and collected the
results at the end, the walkers would be unevenly distributed among
the processes, leading to an inefficient DMC code. For a DMC code to
function properly it needs an as large as possible number of walkers
to get a good representation of the wave function. A lot of
unconnected DMC simulations will basically yield a set of not-so-good
wave functions. We therefore have
to collect all the walkers, remove dead walkers and make copies of the
more virile walkers according to the branching process and then
redistribute the walkers at every time step.

The parallelization is realized by a division of the walker array. A
master node stores an array of the full number of walkers and
distribute these walkers evenly between the slave nodes where the
walkers are stored in smaller walker blocks. The preparation to
sending and receiving the walker blocks is identical to the
checkpoint procedure mentioned above, sans the file writing and
reading. In fact we use the MPI\_Pack procedure to pack the walkers for
checkpointing, even in the serial program. The only difference is that
we send and receive the walkers instead of writing to and reading from
file.

The main problem left is then to ensure that the sets of random
numbers in fact are independent.

\subsubsection{Generating random numbers in parallel}

Generating random numbers in parallel is not as straight-forward as
one may think. A common first approach is to start the same random
generator on every node, varying the seed with the rank of the node as
a factor to get independent streams
and hoping that these streams are uncorrelated. The main problem with
this approach is that random generators often have long-term
correlations which is of little importance in the serial case, but may
appear as short-term correlations in a parallel case
\cite{SPRNG,testingPRNG}. In the extreme case, we may chose seeds
yielding random numbers separated with exactly one cycle. In this case
we will end up with $N_{CPU}$ identical streams, yielding $N_{CPU}$
identical simulations and extremely good (but wrong) statistics in the
results. Several approaches to get safe streams in parallel are
suggested in \cite{SPRNG,testingPRNG} and implemented in the SPRNG
library which we use in our simulations.

\subsection{Python implementation I}

The C++ implementation uses about 90\% of the time in the walker
objects and most of this time in computing local energies ($N^3$
operations where $N$ is the number of particles) in functions located
in Func. In the python implementation of diffusion Monte Carlo (pyDMC)
the classes Walker and Func are therefore linked into a shared library
readable from Python, through a thin wrapper module, together with the
functions from the DMC class below the function oneTimeStep() in
fig. \ref{fig:float_diagram_dmc}.

The main obstacle in implementing pyDMC is the handling of the
walkers. In a straight forward approach we put the walker objects in a
native Python array. This is a very tempting approach; we can leave the
entire problem of creating and killing walkers\footnote{which is not a
straight forward problem with C++ arrays.} to Python. Another approach
is to make a walker array class in C++. This way we can avoid
explicit looping in the Python code, but we are again left to take
care of varying array sizes in C++.

To understand the first approach, we must have a look at how to put the
walkers into a native array. To get a C++ class visible from Python it
has to be compiled and linked into a shared library. This step is
taken care of by the use of SWIG \cite{swig}. 
The walker array is then realized by the function warray, listing
\ref{lst:warray}.
\lstset{language=Python}
\begin{lstlisting}[float=*,caption=warray function,label=lst:warray]
def warray(self,size,particles,dim):
    """function returning an array of (initialized) walkers"""
    w = []
    for i in range(size):
        w += [Walker()]
        w[i].pyInitialize(particles,dim)
    return w
\end{lstlisting}

A great advantage with Python is that you can expand a class in
run-time (or in fact build an entire class in run-time). Utilizing
this advantage, we have inserted the function warray into
the class Walker where it naturally belongs, as can be seen in the
function funcToMethod, listing \ref{lst:funcToMethod}.
\begin{lstlisting}[float=*,caption=funcToMethod,label=lst:funcToMethod]
def funcToMethod(func, clas, method_name=None):
    """function to insert a python func into a class
    Taken from Python Cookbook, recipe 5.12"""
    setattr(clas, method_name or func.__name__, func)
funcToMethod(warray,Walker) #insert function warray in class Walker
\end{lstlisting}
This approach is particularly handy if we want to expand the Func
class with new physical systems, enabling us to write the new functors
in pure Python code. However, as the functors are where most of the
computation time takes place, this approach will severely hinder the
effectiveness of the simulation.

In the native array approach most of the parallelization is realized
with the functions spread\_walkers and gather\_walkers, listing
\ref{lst:spread_n_gather}.
\begin{lstlisting}[float=*,caption=Spread and gather,label=lst:spread_n_gather]
    def spread_walkers(self):
        """Function converting walkers to numpy arrays, sending,
        recieving and converting back"""
        if self.master:
            displace = self.loc_walkers[self.master_rank]
            for i in range(1,self.numproc):
                send_w = self.w[displace:displace+self.loc_walkers[i]]
                send_buff = walkers2py(send_w)
                self.pypar.send(send_buff,i)
                displace += self.loc_walkers[i]
        else:
            recv_buff = self.pypar.receive(self.master_rank)
            w_args = [self.loc_walkers[self.myrank], self.particles,\
                      self.dimensions]
            self.w_block = py2walkers(recv_buff, *w_args)


    def gather_walkers(self):
        """Look at spread_walkers in a mirror"""
        if self.master:
            displace = self.loc_walkers[self.master_rank]
            for i in range(1,self.numproc):
                recv_buff = self.pypar.receive(i)
                w_args = [self.loc_walkers[i], self.particles,\
                          self.dimensions]
                self.w[displace:displace+self.loc_walkers[i]] = py2walkers(\
                    recv_buff, *w_args)
                displace += self.loc_walkers[i]
        else:
            send_buff = walkers2py(self.w_block)
            self.pypar.send(send_buff,self.master_rank)
\end{lstlisting}
The functions walkers2py and py2walkers are functions for
converting a walker to a NumPy array and back again, taking advantage
of the functions getBuffer and setBuffer in the Walker class.

An example of a parallelized diffusion Monte Carlo program is 
realized in less than 100 lines:
\begin{widetext}
\lstset{basicstyle=\tiny}
\begin{lstlisting}
    >>> import pypar,math
    >>> from DMC import DMC
    >>>
    >>> d = DMC(pypar)
    >>> # 1 particle and alpha=0.5 yields a harmonic oscillator with
    >>> # energy == 3/2:
    >>> d.params[0] = 0.5
    >>> d.reset_params()
    >>> d.silent = True # to avoid too much noise
    >>>
    >>> def timestep(i_step):
    ...     M = d.no_of_walkers
    ...     d.spread_walkers()
    ...     for walker in d.w_block:
    ...         d.monte_carlo_step(walker)
    ...     d.gather_walkers()
    ...     d.update = False
    ...     if d.master:
    ...         #bring out your dead
    ...         for i in range(M-1,-1,-1):
    ...             if d.w[i].isDead():
    ...                 d.w[i:i+1] = [] #removing walker
    ...             else:
    ...                 while d.w[i].tooAlive():
    ...                     baby_walker = d.copy_walker(d.w[i])
    ...                     baby_walker.calmWalker()
    ...                     d.w += [baby_walker]
    ...                     d.w[i].madeWalker()
    ...         d.no_of_walkers = len(d.w)
    ...     d.no_of_walkers = d.pypar.broadcast(d.no_of_walkers,
    ...                                         d.master_rank)
    ...     d.refresh_w_blocks()
    ...     d.spread_walkers()
    ...     d.num_args[-1] = d.update
    ...     if d.master:
    ...         nrg = 0.; pot_nrg = 0.; vort_nrg = 0
    ...         d.time_step_counter += 1
    ...         for walker in d.w:
    ...             d.num_args[-1] = d.update
    ...             nrg       += walker.getLocalEnergy(*d.num_args)
    ...             pot_nrg   += walker.getLocalEnergy(d.pot_e)
    ...             if hasattr(d,'vort_e'):
    ...                 vort_nrg += walker.getLocalEnergy(vort_e)
    ...         nrg      /= float(d.no_of_walkers*d.particles*d.scale)
    ...         pot_nrg  /= float(d.no_of_walkers*d.particles*d.scale)
    ...         vort_nrg /= float(d.no_of_walkers*d.particles*d.scale)
    ...         d.energy  += nrg
    ...         d.energy2 += nrg*nrg
    ...         if d.time_step_counter >= d.termalization:
    ...             #implement pos_hist to be called here
    ...             d.observables[i_step,0] = nrg
    ...             d.observables[i_step,1] = pot_nrg
    ...             d.observables[i_step,2] = vort_nrg
    ...         # adjust trial energy (and no. of walkers)
    ...         nrg = -.5*math.log(float(d.no_of_walkers)/float(M))/d.tau
    ...         d.e_trial += nrg
    ...     d.e_trial = d.pypar.broadcast(d.e_trial,d.master_rank)
    ...     d.no_of_walkers = d.pypar.broadcast(d.no_of_walkers,
    ...                                         d.master_rank)
    ...
    >>> #set initial walker positions:
    >>> if d.metropolis_termalization: d.uni_dist()
    ...
    >>> for i in range(d.metropolis_termalization):
    ...     for walker in d.w_block:
    ...         d.metropolis_step(walker)
    ...
    >>>
    >>> nrg = 0.
    >>>
    >>> for walker in d.w_block:
    ...     nrg += walker.getLocalEnergy(*d.num_args)
    ...
    >>> d.gather_walkers()
    >>> d.e_trial = d.all_reduce(nrg)
    >>> d.time_step_counter = 0
    >>> for i in range(d.termalization):
    ...     timestep(i)
    ...
    >>> d.energy = 0; d.energy2 = 0
    >>> for i in range(d.steps):
    ...     timestep(i)
    ...
    >>> if d.master:
    ...     d.energy  /= float(d.steps)
    ...     d.energy2 /= float(d.steps)
    ...     d.energy2 -= d.energy*d.energy
    ...     print "energy = %g +/- %g"%(d.energy,
    ...                                 math.sqrt(d.energy2/float(d.steps)))
    ...     print "sigma = %g"%d.energy2
    ...
    energy = 1.5 +/- 0
    sigma = 0
    >>> pypar.Finalize()
\end{lstlisting}
\lstset{basicstyle=\small}
\end{widetext}

\subsection{Python implementation II}

When thinking performance of arrays in Python the add-on package 
{\it Numerical Python} (NumPy) springs to mind as an obvious
choice. The fact that the module pypar (which we are going to use in
the parallelization) supports sending NumPy arrays directly, is of
course helping in that choice. However, even though NumPy supports a
lot of types, (such as integers, floats, chars etc.) there is no
support for walkers as a type\footnote{To the best of my
  knowledge}. It is possible to use generic python objects in NumPy
arrays, but this will mainly make NumPy array comparable to native arrays.

The approach with native python arrays is quite straight-forward and
easy to implement. It is, however, quite inefficient as well. There are
two main reasons for this. First, looping is known to be an
inefficient construct in Python. With a native Python array, the loops
over walkers have to be done in Python. Second, native python arrays
are slower than, e.g., NumPy arrays, because native arrays are written
for a much more general use than just numerics. The question is how we
can use the most of the C++ walker code as-is while avoiding explicit
for-loops in Python.

One solution is to implement an array class (lets call it
WalkerArray) in Python which is wrapper class to a C++ class
containing a C++ array of walkers and functionality to create and kill
walkers. Even though this is a good approach in a serial
implementation, we still have to convert these arrays to NumPy arrays
to be able to send the walkers in MPI.

In our Python implementation, we keep the WalkerArray, but
store all the walker data in a NumPy array. To do this we have
modified the C++ Walker class so that it only uses pointers to an
array for all 
arrays and variables that should be stored. This array is then
provided by NumPy. Even though this approach taints the C++
implementation of the Walker class, we get the advantage that the
Python DMC class only has to care about NumPy arrays, providing us
with powerful tools for vectorizing the Python code.

\subsubsection{WalkerArray}

To implement the class WalkerArray we need to have some knowledge of
the C++ Walker class. The Walker class contains some arrays storing
all particle positions and all previous particle positions and variables
to know if the walker should be removed or duplicated. In
addition it stores the last computed local energy, quantum force and
wave function to minimize the number of times we compute these
quantities. In the C++ code this information is allocated and stored in each
walker, making the creation of a walker rather costly. When we send
walkers in MPI the information is collected from each 
walker and concatenated into one array before communication and
inserted into the walkers after communication. Now, we want turn it
the other way, i.e., we want to allocate and store all information from all
walkers in one array (preferably a NumPy array) and let the walkers
operate on pointers to this array. This way the time to initialize a
walker is reduced dramatically and all information on walkers are
readily available from Python. This change of view for the Walker
class is realized by changing all variables that define the walker to
references to the corresponding pointers to the NumPy array.

\section{Python vs. C++}
\label{sect:benchmarks}

Now we know how to implement a Python version of a Monte Carlo
solver. We then need to know if MontePython is efficient enough. We
can assume that the Python implementation will never be faster that
the corresponding C++ code, as Python will always have some degree of
overhead just to access the C++ code. The question is how big this
overhead may be. 
\begin{figure*}[t]
  \begin{center}
    \includegraphics[width=0.45\textwidth]{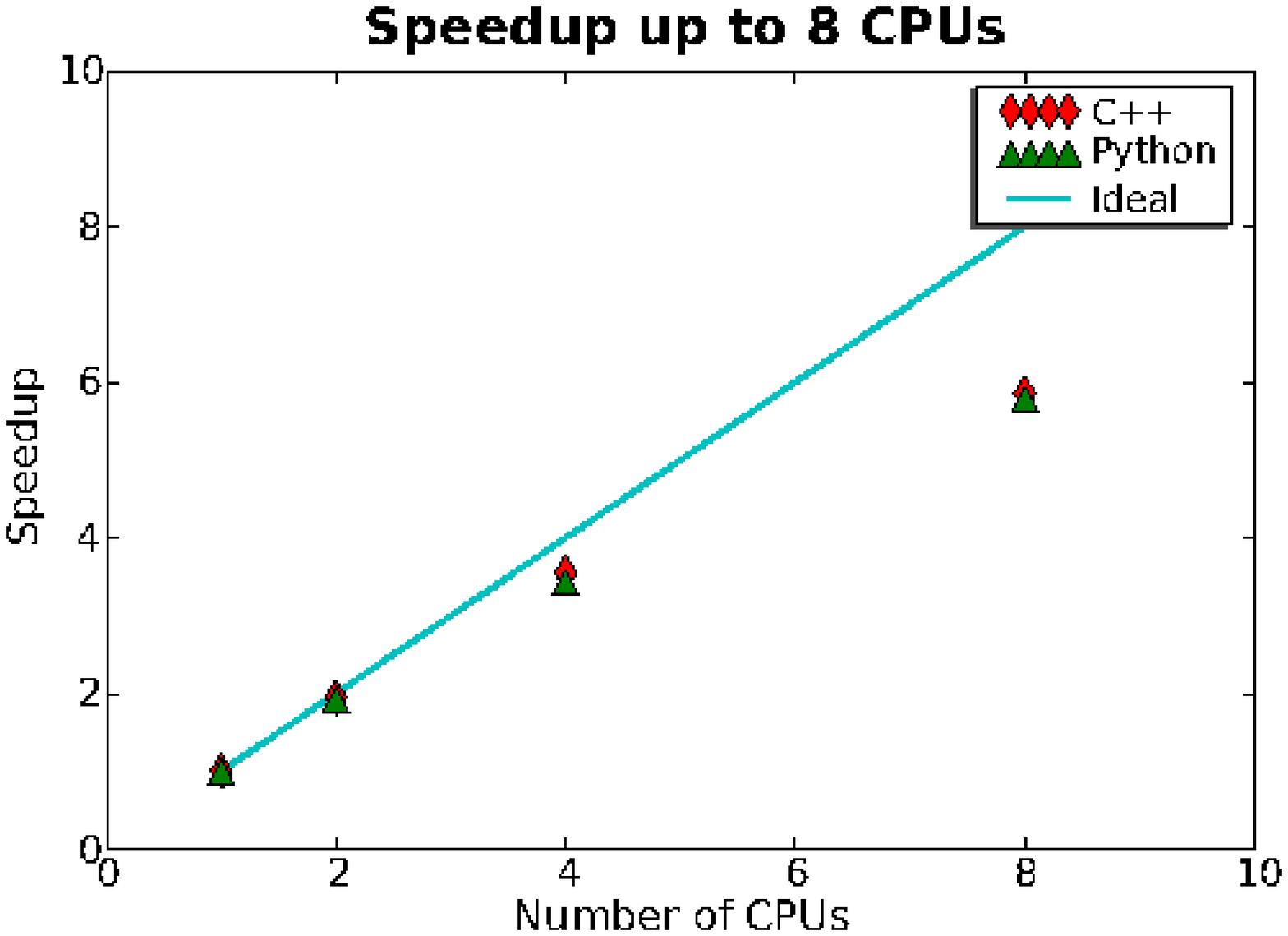}
    \includegraphics[width=0.45\textwidth]{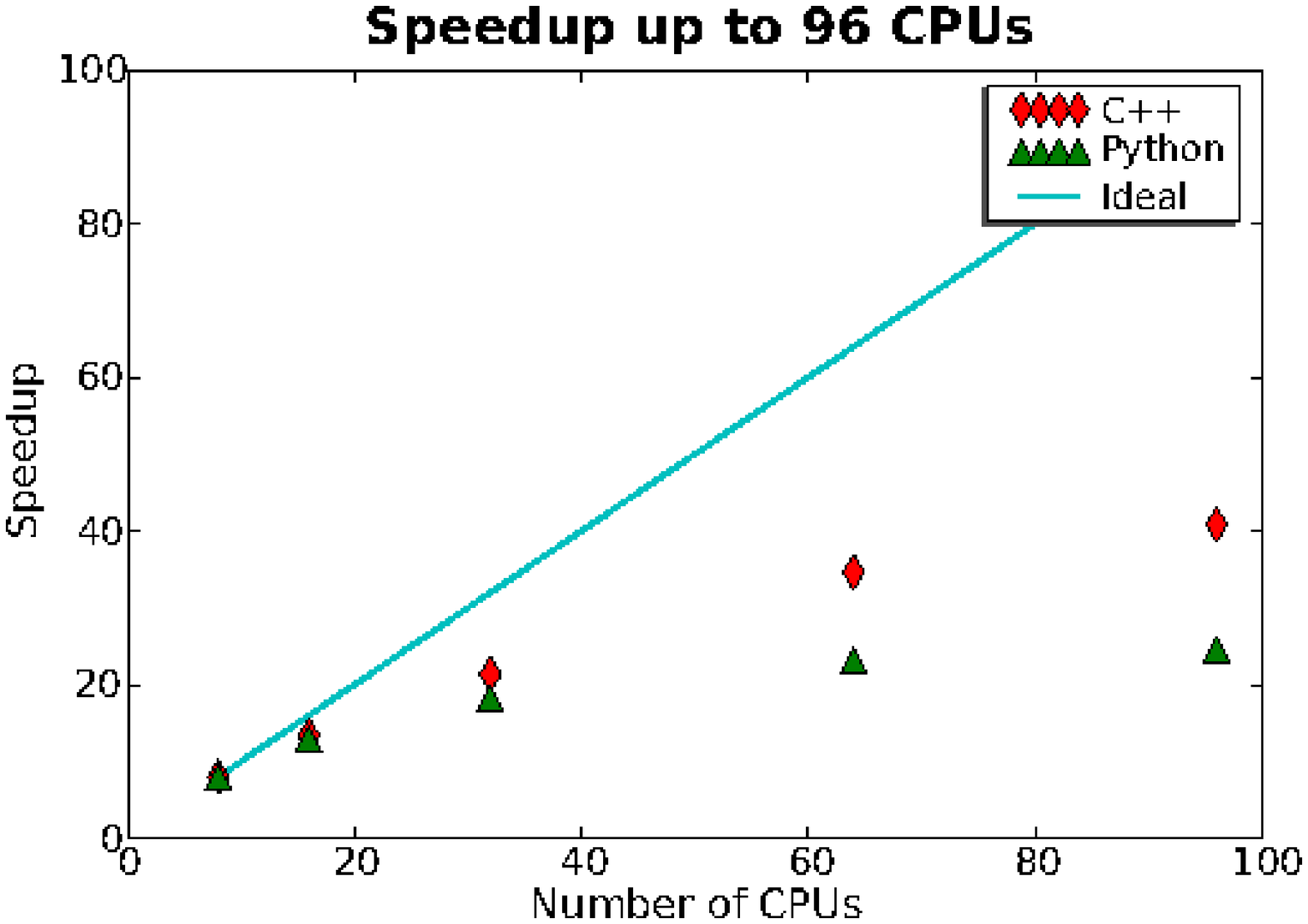}
  \end{center}
  \caption{Speedup of a simulation as a
  function of the number of CPUs used. In the left figure the serial
  run took about 30 minutes and was run with an initial 480 walkers
  moved in 3500 time steps. In the right figure the serial run
  took about 4 hours 30 minutes and was run with an initial 4800
  walkers moved in 1750 time steps.} 
  \label{fig:dmc_python_speedup}
\end{figure*}
In figure \ref{fig:dmc_python_speedup} we have
plotted the speedup of a Monte Carlo simulation as a function of the
number of CPUs\footnote{Speedup is the time of a serial simulation
divided by the walltime of the simulation.}. To the left we have done a
relatively light simulation 
with 20 particles in 3 dimensions using 480 walkers. The walkers are
spread out evenly and communicated from the master node to the slave
nodes and back again 3500 times so that all walkers are sent 7000
times. The size of one walker is in this case $1.2$kB which means that
for, e.g., 4 CPUs the size of each message is about 144kB. We can see
that for this message size the overhead of using Python is almost
none.

To the right in figure \ref{fig:dmc_python_speedup} we have increased
the number of walkers to 4800. However, we are also using a much higher
number of CPUs. The message size for, e.g., 64 CPUs is only 9kB. As
we explicitly loop over the number of CPUs when we send walkers to
and from the master, the overhead increases dramatically when compared
to the time to send one message. The send and
receive methods should therefore be vectorized with scatter and gather
routines. Unfortunately we do not have uniform message sizes, making
generic scatter and gather routines unusable. We will therefore have
to write these routines ourselves.

Python has similar speedup
to C++, but the curve flattens out much faster as we increase the
number of CPUs. As Monte Carlo simulations are known to
have perfect speedup, we cannot be satisfied with the parallel
algorithms of either the C++ version or the Python version.

\section{Optimizing MontePython}

One of the points in using Python in scientific programming is that
you can implement new and improved algorithms efficiently. We have
seen that distribution of the random walkers over the compute nodes
leads to a bottleneck due to communication when the number of CPUs
grows large. This bottleneck is evident both for the C++
implementation and the Python implementation. In this
section we will improve the algorithm for load balancing of the
walker in two ways. First, we will improve on the way the walkers are
killed and reproduced. Second, we will improve on the load balancing
itself by optimizing for heterogeneous clusters of CPUs.

\subsection{Distributing Walkers in Parallel}

The C++ Diffusion Monte Carlo application was originally written in
serial and then ported to parallel using MPI. In the serial version we
used an algorithm where, in order to kill a walker, we moved the last
walker in the sequence onto the walker that was to be killed and
decreased the number of walkers with one. To reproduce a walker, we
copied the walker to the end of the sequence. This algorithm is
optimal and widely used in serial Diffusion Monte Carlo. When we
parallelized the code, we kept this serial algorithm by gathering all
the walkers to the master node, let the master node do the killing and
reproducing in serial, and then spread the walkers evenly among the slave
nodes again. This way the load was always balanced, and the master had
full control of the walkers at all times. The main problem is that,
apart from memory issues as the master needs to store a lot of walkers,
the serial work load for the master increases fast when we increase
the number of CPUs in the calculation. This problem is very clear
in figure \ref{fig:dmc_python_speedup}, where the speedup is quite
poor already for 32 CPUs, both for the C++ version and the Python
version.

Again, the algorithm is best explained through the source code. First
we let the slave nodes individually move their walkers and kill and
reproduce their local walkers, then the function
\texttt{DMC.load\_balancing()},balances the load:
\lstset{language=Python}
\lstset{numbers=left, numberstyle=\tiny, stepnumber=2, numbersep=5pt}
\lstset{escapeinside={(*@}{@*)}}
\begin{widetext}
\begin{lstlisting}
    def load_balancing(self):
        """Function balancing load between nodes"""
        self.t1 = time.time()
        w_numbers = self.pypar.gather(Numeric.array([len(self.w_block)]),
                                      self.master_rank)(*@\label{lst:gather_w_num}@*)
        tmp_w_numbers = copy.deepcopy(w_numbers)
        w_numbers = self.pypar.broadcast(tmp_w_numbers,
	                                 self.master_rank)(*@\label{lst:bcast_w_num}@*)

        self.no_of_walkers = Numeric.sum(w_numbers)(*@\label{lst:no_of_walkers}@*)

        self.__find_opt_w_p_node()(*@\label{lst:opt_walkers}@*)

        self.first_balance = False
        balanced = Numeric.array(self.loc_walkers)

        difference = w_numbers-balanced(*@\label{lst:difference}@*)

        diff_sort = Numeric.argsort(difference)
        prev_i_min = diff_sort[0]
        
        while sum(abs(difference))!=0:(*@\label{lst:sum_abs_diff}@*)
            diff_sort = Numeric.argsort(difference)
            i_max = diff_sort[-1]
            i_min = diff_sort[0]
            
            if i_min == prev_i_min:(*@\label{lst:skip_busy_node}@*)
                i_min = diff_sort[1]
            
            if self.myrank==i_max:(*@\label{lst:start_send_recv}@*)
                self.pypar.send(self.w_block[balanced[i_max]:],i_min)
                args = [balanced[i_max],
                        self.particles,
                        self.dimensions,
                        self.w_block[0:balanced[i_max]]]
                self.w_block = WalkerArray.WalkerArray(*args)
            elif self.myrank==i_min:
                recv_buff = self.pypar.receive(i_max)
                args = [len(self.w_block)+difference[i_max],
                        self.particles,
                        self.dimensions,
                        Numeric.concatenate((self.w_block[:],recv_buff))]
                self.w_block = WalkerArray.WalkerArray(*args)(*@\label{lst:stop_send_recv}@*)
            difference[i_min]+=difference[i_max](*@\label{lst:update_diff_min}@*)
            difference[i_max]=0(*@\label{lst:update_diff_max}@*)
            prev_i_min = i_min
\end{lstlisting}
\end{widetext}

This function deserves some explanation. From line
\ref{lst:gather_w_num} to \ref{lst:no_of_walkers} we update the
current walker distribution and total number of walkers. In line
\ref{lst:opt_walkers} we determine the optimal distribution of
walkers. to be explained in subsection
\ref{sect:heterogeneous_clusters}. At this point we know the actual
distribution of walkers and the optimal distribution of walkers. The
idea is then to find the length of the difference between the optimal
and actual distribution and move walkers among nodes until the
length, or the sum of the absolute value of the differences is 
is zero, see line \ref{lst:sum_abs_diff}. This is realized in
lines \ref{lst:start_send_recv}-\ref{lst:update_diff_max} by moving
the excess walkers from the node with maximum difference to the walker
with minimum difference recursively. A problem with this procedure is that
the the same node can have a minimum difference in subsequent cycles
of the while-loop\footnote{E.g., if a node with minimum difference
  needs 4 walkers and the second minimum is 1, while the maximum
  difference is 2, the minimum node is still the same after the first
  cycle. Then the message of the next cycle will have to wait till the
  first message is sent.}. This leads to unnecessary waiting in the
program. The remedy is seen in line \ref{lst:skip_busy_node} where we
take the second minimum node if the minimum node is the same node as
in the previous cycle. Of course this problem may just be transferred
to the second minimum node, but this is much less likely to happen. 

It should be noted that this optimization does not preserve the result
from the non-optimized code, in the sense that we will not get an identical
output in the end. This is due to the fact that the random sequences
are distributed per node and not per walker, meaning that each walker
will get a different series of random numbers depending on which node
it is sent to. The output will, however, be within the error range of
the non-optimized code.

\subsection{Heterogeneous clusters}
\label{sect:heterogeneous_clusters}

Most new high performance clusters are more or less homogeneous,
in the sense that the computation nodes have identical specifications
with respect to CPU, RAM, network and storage. However, as a cluster
usually expands in time, due to more funds and need for more
resources, it is very likely that it will become a heterogeneous
cluster\footnote{Not only because the funders always want bleeding edge
  technology, but simply because computation nodes with the old
  specifications are not produced any longer.}. Also, with the trend of
multiple cores and CPUs per computation nodes, combined with the fact
that there is more than one user per cluster, different nodes will
have different (and possibly too high) load and therefore different
computational speed. This means that even if a cluster is homogeneous
on paper, it will act like a heterogeneous cluster in practice. If we
want to gain the optimal performance from a cluster, we need to take
into account this heterogeneity.

In the function \texttt{\_\_find\_opt\_w\_p\_node()} we use the time
from the DMC class is initialized to the function is called to
determine how the optimal distribution of walkers at every time
step. The function itself can be seen in listing \ref{lst:opt_w_p_node}.
\begin{lstlisting}[float=*,caption=Finding optimal walkers per node,label=lst:opt_w_p_node]
     def __find_opt_w_p_node(self):
        """Help function for load_balancing()
        Finds and returns the optimal number of walkers
        per node for a possibly non-uniform set of nodes
        """
	self.t1 = time.time()
        timings = self.pypar.gather(Numeric.array([abs(self.t1-self.t0)]),
                                    self.master_rank)
        tmp_timings = copy.deepcopy(timings)
        timings = self.pypar.broadcast(tmp_timings,
                                       self.master_rank)

        C = self.no_of_walkers/sum(1./timings)(*@\label{lst:time_constant}@*)

        tmp_loc_walkers = C/timings

        self.loc_walkers = self.NumericFloat2IntList(tmp_loc_walkers)
        remainders = tmp_loc_walkers-self.loc_walkers

        while sum(self.loc_walkers) < self.no_of_walkers:
            maxarg = Numeric.argmax(remainders)
            self.loc_walkers[maxarg] += 1
            remainders[maxarg] = 0

        if self.master and self.first_balance:
            print timings
            print self.loc_walkers

        return self.loc_walkers
\end{lstlisting}
To understand this function we just need some simple linear
algebra. Say that we have set of walkers \matenv{[x_1,x_2,\ldots x_N]}
spread in an optimal way over $N$ nodes. On node $i$ the time to move
one walker is given by $a_i$ yielding a set \matenv{[a_1,a_2,\ldots
    a_N]}. By {\it optimal distribution} we mean a distribution of 
walkers were each node finishes the work assigned to it between
synchronizations at the same time $C$, i.e. \matenv{a_ix_i=C}. In
addition we know the total number of walkers, $T$. We know that
\begin{equation}
\sum x_i = \sum\frac{C}{a_i},
\end{equation}
so that the problem reduces to finding $C$. However, as $T=\sum x_i$, we have
\begin{equation}
C=\frac{T}{\sum 1/a_i}.
\label{eq:time_constant}
\end{equation}
The zealous reader will notice that eq. \refeq{eq:time_constant}
corresponds to line \ref{lst:time_constant} of 
\texttt{\_\_find\_opt\_w\_p\_node()}. The rest of the function is
merely taking care of the fact that $x_i$ are integers
while $a_i$ and $C$ are real numbers.

%%%%%%%%%%%%%%%%%%%%%%%results%%%%%%%%%%%%%%%%%%%%%%%%%%%%%%%%%
\subsection{Optimized results}

\begin{figure*}[t]
  \begin{center}
    \includegraphics[width=0.45\textwidth]{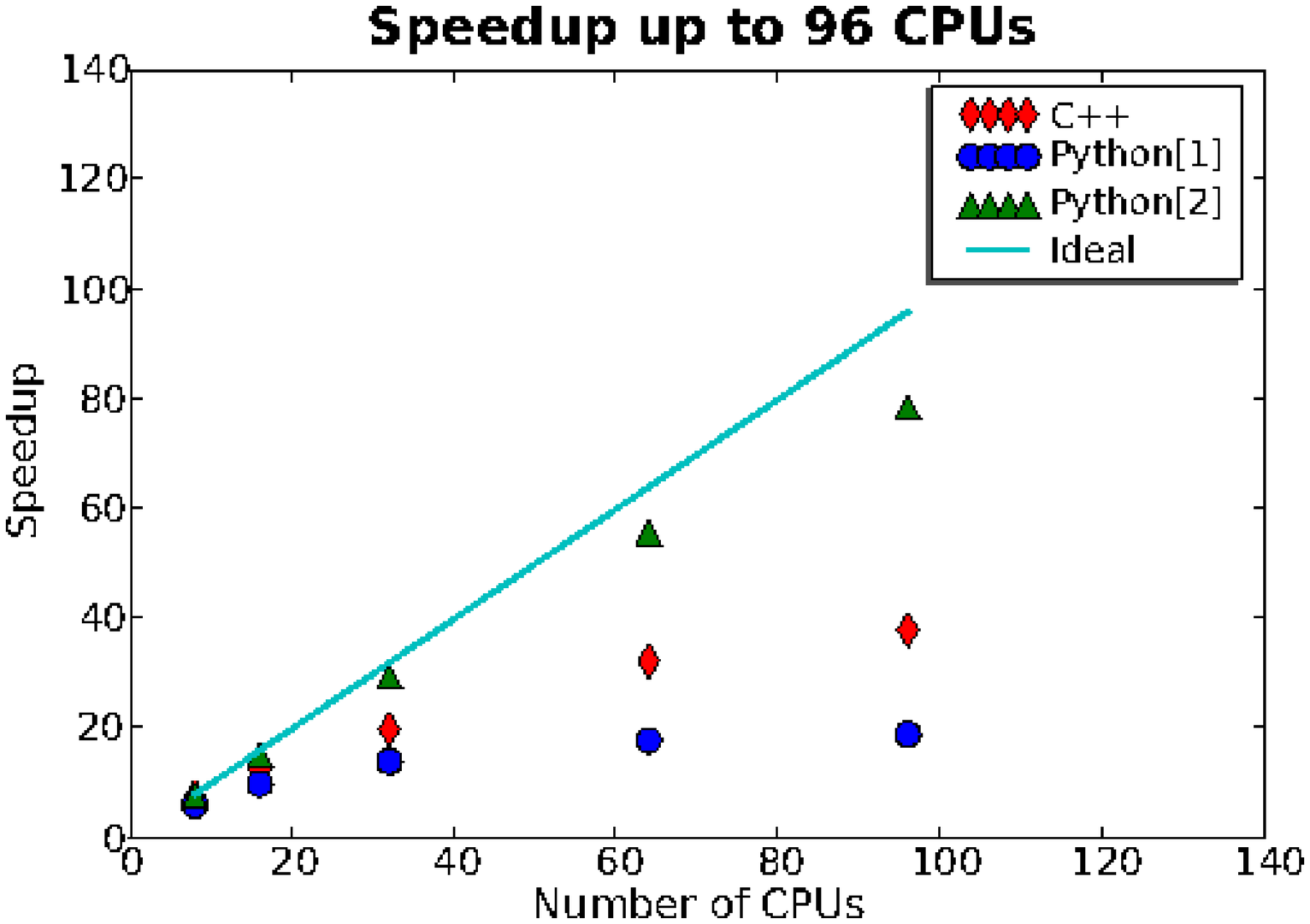}
    \includegraphics[width=0.45\textwidth]{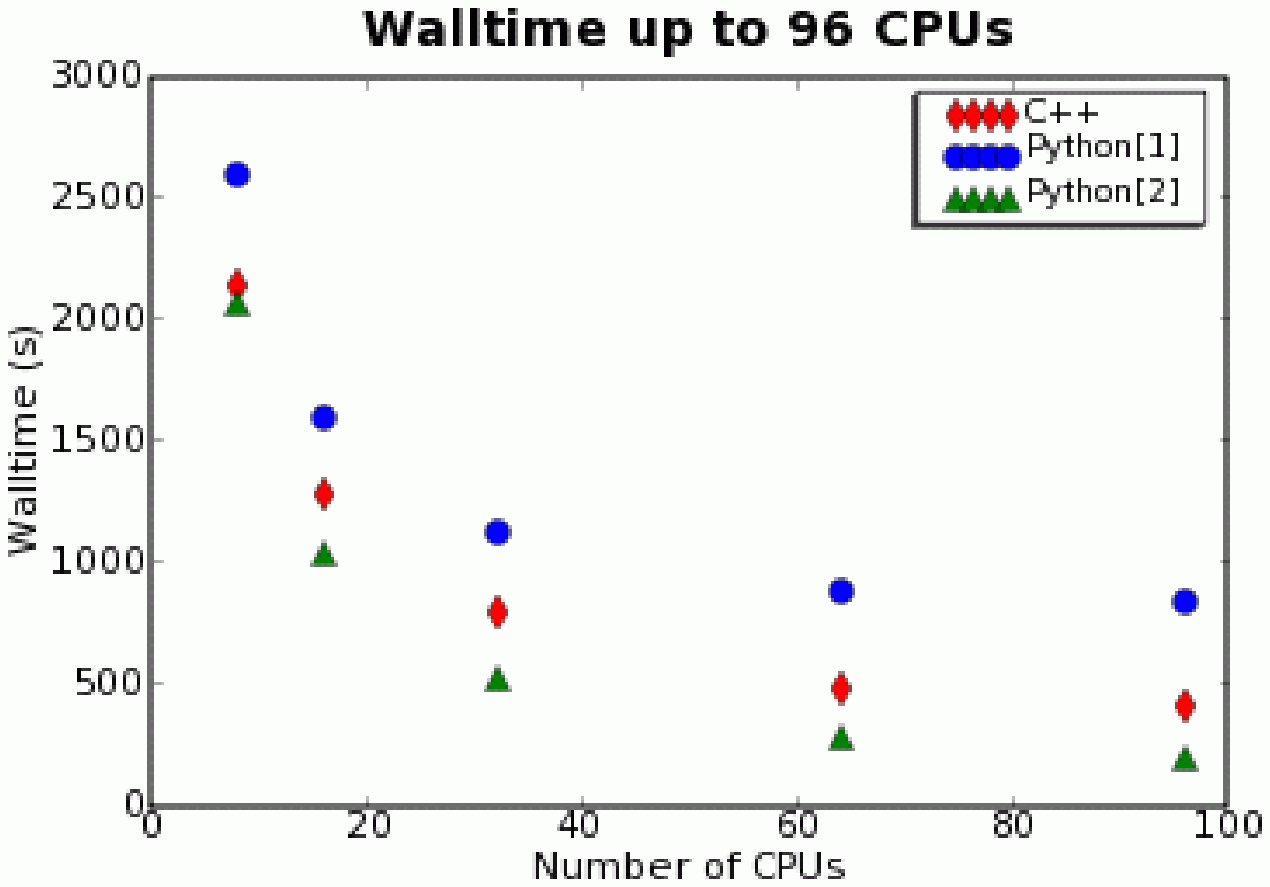}
  \end{center}
  \caption{The figures show the speedup (left figure) and walltime
  (right figure) for the simulation as a 
  function of the number of CPUs used. The serial
  run took about 270 minutes and was run with an initial 4800 walkers
  moved in 1750 time steps.} 
  \label{fig:dmc_python_optpython_speedup_walltime}
\end{figure*}

In figure \ref{fig:dmc_python_optpython_speedup_walltime} we show the
speedup and walltime of the improved walker distribution compared to
the non-optimized C++ and Python version from figure
\ref{fig:dmc_python_speedup}. We see
that both the speedup and the walltime is much more reasonable for the
optimized version of MontePython. In fact, the new version is more
than twice as fast as the C++ version for 96 CPUs.

These optimizations would of course be possible to do in the C++
application as well, albeit in more than 75 lines. However, the simple
syntax in Python and the use of Numeric arrays to store walkers allow
us to concentrate our effort directly on the optimization of the
algorithm instead of dealing with, e.g., how to send, receive and
concatenate a slice of walkers.

\section{Visualizing with Python}
\label{sect:visualizing}

\begin{figure}[t]
  \begin{center}
    \includegraphics[width=0.5\columnwidth]{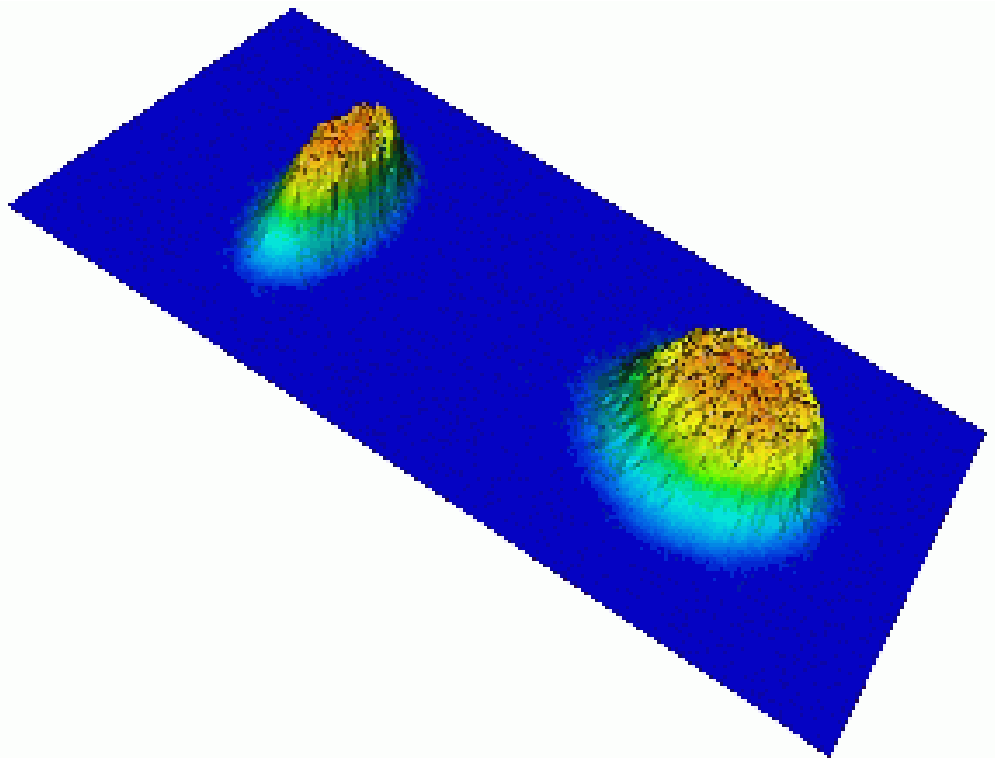}\\
    \includegraphics[width=0.5\columnwidth]{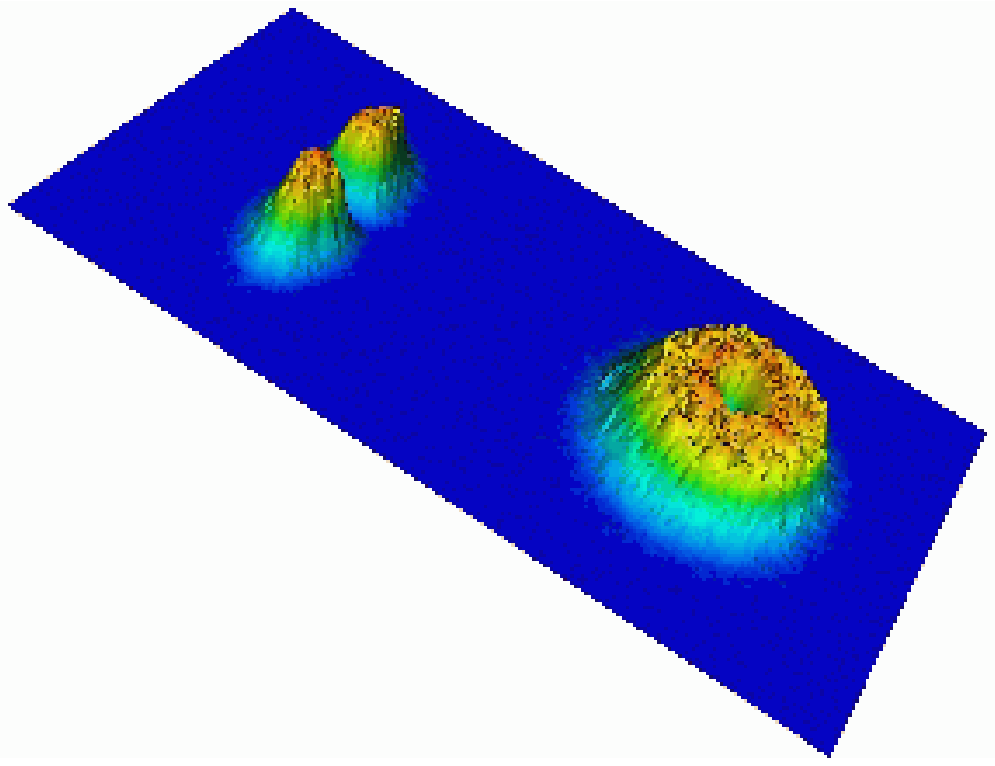}
  \end{center}
  \caption{The figures show where particles are detected. We see the
  expectation values in two spatial dimensions, the $yz$-plane to the
  left and the $xy$-plane to the right. The topmost figure corresponds
  to the ground state, while the bottom figure corresponds to a state
  with one vortex in the center of the trap.} 
  \label{fig:expectation2d}
\end{figure}

\begin{figure}[t]
  \begin{center}
    \includegraphics[width=0.8\columnwidth]{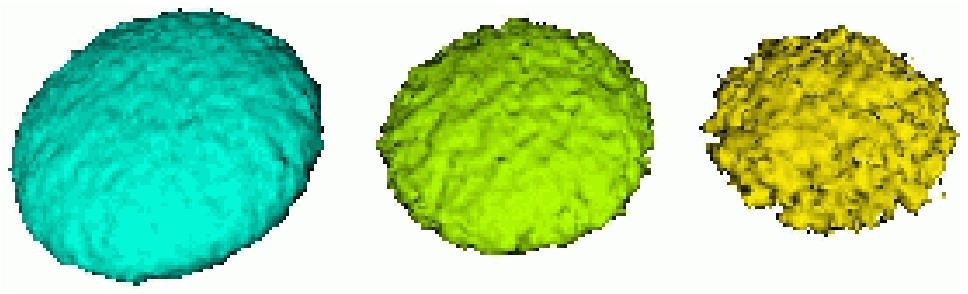}
    \includegraphics[width=0.8\columnwidth]{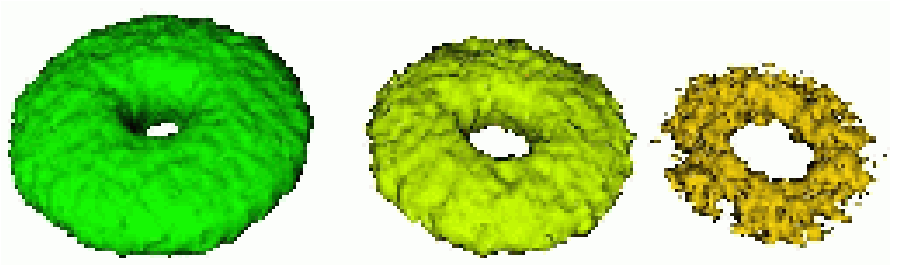}
  \end{center}
  \caption{The figures show where particles are detected. We see the
  expectation values for finding, from left to right, 1, 2, 3 and 4
  particles. The topmost figure corresponds
  to the ground state, while the bottom figure corresponds to a state
  with one vortex in the center of the trap.} 
  \label{fig:expectation3d}
\end{figure}

We mentioned in the introduction that integration of simulation
and visualization is an important feature of Matlab, Maple and
others. This feature is maybe even more powerful in Python. In figures
\ref{fig:expectation2d}, \ref{fig:expectation3d} and
\ref{fig:expectation3d} we have used pyVTK
and Mayavi \cite{mayavi} to plot the particle density, which is an output of
diffusion Monte Carlo. pyVTK \cite{pyvtk} is a python interface to the
Visualization ToolKit (VTK), while Mayavi, which is built
on pyVTK, is
a scriptable graphic interface for 3D visualization.

A signature of a Bose-Einstein condensate is that it is
irrotational. If we try to rotate the condensate, it will compensate by
setting up quantum vortices along the rotational axis. Vortices is
therefore crucial to the study of Bose-Einstein condensates. We need
only small modifications to the Hamiltonian (eq. \refeq{eq:Hamiltonian})
and trial wave function (eq. \refeq{eq:wf}) to consider a single vortex
along the $z$-axis in our system \cite{nilsen2005}.

Figures \ref{fig:expectation2d} and \ref{fig:expectation3d} shows the
change in the ground state when inserting the vortex. The repulsive
nature of the vortex pushes the particles away from the $z$-axis,
decreasing the maximum density when compared to the ground
state.

\section{Conclusion}
\label{sect:conclusion}

We have implemented a Monte Carlo solver using three different
approaches, from pure C++, through a straight forward Python
implementation, to an efficient, vectorized Python
implementation. Furthermore we have compared the C++ and the
vectorized Python implementations and shown that the overhead of using
Python is non-existent for sufficiently large problems. In fact, with
only 75 lines of code we were able to introduce an improved parallel
algorithm for walker distribution, to gain a super-linear speedup when
compared to C++. In addition, we have shown that Python can be used
directly as a visualization tool for rendering three dimensionally
scientific visualizations. We can therefore safely say that Python
serves as a powerful tool in scientific programming.

\newpage
\bibliography{jon_nilsen}
\bibliographystyle{unsrt}

\end{document}